\documentclass{elsart}

\usepackage{epsfig}
\usepackage{cite}

\usepackage{amssymb}

\def\be{\begin{equation} }
\def\ee{\end{equation} }
\def\ba{\begin{eqnarray} }
\def\ea{\end{eqnarray} }
\def\ban{\begin{eqnarray*} }
\def\ean{\end{eqnarray*} }

\def\arco{\mbox{Ar:CO$_2$}}

\def\ExB{\mbox{\boldmath$\rm E \times B$\unboldmath}}
\def\mum{\mbox{$\mu$m}}

\def\FWHM{\mbox{\it FWHM}}

\begin{document}

\begin{frontmatter}


\title{Spatial resolution of a GEM readout TPC using
the charge dispersion signal}

\author[ref_CU]{K.~Boudjemline},
\author[ref_CU,ref_TR]{M.~S.~Dixit\corauthref{add_MD}},
\ead{<msd@physics.carleton.ca>}
\author[ref_UM]{J.-P.~Martin}
and
\author[ref_CU]{K.~Sachs}

\address[ref_CU]{Department of Physics, Carleton University, 
        \\ 1125 Colonel By Drive, Ottawa, ON, K1S 5B6, Canada}
\address[ref_UM]{Universit\'e de Montr\'eal, Montr\'eal, QC H3C 3J7, Canada}
\address[ref_TR]{TRIUMF, Vancouver, BC V6T 2A3, Canada}
\corauth[add_MD]{Corresponding author; 
         tel.: +1-613-520-2600, ext. 7535; fax: +1-613-520-7546.}


\begin{abstract}
A large volume Time Projection Chamber (TPC) is being considered for the central charged particle tracker for the  detector for the proposed International Linear Collider (ILC).  To meet the ILC-TPC spatial resolution challenge of $\sim$ 100 $\mum$ with a manageable number of readout pads and channels of electronics,  Micro Pattern Gas Detectors (MPGD) are being developed which could use pads comparable in width to the 
proportional-wire/cathode-pad TPC.   We have built a prototype GEM readout TPC with 2 mm x 6 mm pads using the new concept of charge dispersion in MPGDs with a resistive anode. The dependence of transverse  resolution on the drift distance has been measured for small angle tracks in cosmic ray tests without a magnetic field for \arco (90:10). 
The GEM-TPC resolution with charge dispersion readout is significantly better than  previous measurements carried out with conventional direct charge readout techniques.
\setlength{\unitlength}{1mm}
 \begin{picture}(0,0)
 \put(20,160){\parbox{5cm}{Carleton Phys/061020 }}
\end{picture}
\end{abstract}

\end{frontmatter}

\section{Introduction}
\label{sec:intro}

Large volume time projection chambers (TPC) \cite{cit:TPC1,cit:TPC2} 
have been used as high precision tracking detectors in many high energy
physics experiments since the 1970s. A large volume TPC is also a prime
track detector candidate for future experiments at the International
Linear collider (ILC). However for the ILC application, it will be
important to improve the spatial resolution capability for the TPC.
A promising possibility is the replacement of the traditional proportional-wire/cathode-pad
readout by a Micro Pattern Gas Detector (MPGD) like the Gas Electron Multiplier
(GEM) \cite{cit:gem,cit:gem2}. This would eliminate one of the major 
systematic errors which results from the so called \ExB\ effect 
\cite{cit:ExB} that degrades the TPC spatial resolution.

The readout of a TPC with MPGD has several advantages but also some
drawbacks, both of which are related to the confinement of the signal
charge to a small spatial region due to reduced transverse in a high magnetic field. The advantage is that the localization has the potential to improve the double track resolution. 
The disadvantage with conventional MPGD direct charge readout technique is that it leads to difficulties
with the determination of the signal position. For a nominal
pad size of $\sim$2 mm,  signals may often be confined to one or two pads only
making a centroid calculation less precise, in contrast to the proportional-wire/cathode-pad readout. A smaller pad width would
lead to a better resolution but also to a large number of readout 
channels which may be difficult to manage for a large detector.  

One possibility to improve the signal centroid determination and thus achieve good
resolution with relatively wide pads is to use a MPGD with a resistive anode which
disperses the avalanche charge and allows the signal to be
reconstructed on several pads. The principle of charge dispersion has 
been proven previously \cite{cit:dispersion} for a  GEM using point  
X-ray source. The charge dispersion phenomenon and its application to MPGD-TPC readout are now well understood as shown in the excellent agreement of model simulations with experimental data \cite{cit:simulation}.
In this paper, we present  our first results of MPGD-TPC track resolution measurements with charge dispersion 
for cosmic-ray particles.
The spatial resolution of a GEM-TPC  is measured as a function of drift distance using the charge dispersion technique.  \arco (90:10) was used as a fill gas to mimic the reduced transverse diffusion 
for a TPC  in a high magnetic field.
The results are compared to our previous measurements of 
GEM-TPC resolution \cite{cit:tpc1b}  with direct charge readout for the same gas.

\section{Experimental setup}
\label{sec:exp}

A small 15 cm drift length double-GEM TPC used earlier for cosmic ray resolution studies with conventional
direct charge readout  \cite{cit:tpc1b} was modified for the measurements reported here. The standard anode pad readout plane was replaced with  a resistive anode readout  structure  \cite{cit:dispersion}. 
The new anode structure is fabricated by laminating a 25~\mum\ thick film 
of carbon loaded Kapton with an effective surface resistivity of 
530~${\rm K}\Omega/\square$ to the readout pad PCB using a double 
sided insulating adhesive.
The adhesive provides a 50 \mum\  gap between the resistive anode 
and the PCB.  Taking the dielectric constant of the glue into account, 
the gap results in a capacitance density of C $\simeq$ 0.22 pF/mm$^2$. 
The film surface resistivity and the capacitance  results in a 
$RC$ coupling of the anode to the readout plane.  An avalanche charge 
arriving at the anode surface disperses with the system $RC$ time constant. 
Signals are induced on readout pads as explained in reference \cite{cit:dispersion}.

 The TPC drift field for  \arco (90:10) at 300 V/cm was larger than in our previous measurements with direct charge readout for the same gas. From Magboltz \cite{cit:Magboltz}, we find that the larger drift field increased the electron drift
velocity from 8.9 to 22.75 $\mum/ns$, and decreased the transverse diffusion slightly from
0.229 to 0.223 $\mum/\sqrt{cm}$. Within measurement errors, the effective gas gain for the two measurements was about the same, about 6700.

The layout of the readout pad system contained 5 inner rows of 12 pads each 
(pad size: 2 mm $\times$ 6 mm), together with two large area 
outer pads whose signals are used for selecting cosmic events for analysis. 
Charge signals on the central 60 pads used for tracking were read out using Aleph proportional wire TPC readout
preamplifiers.  We used 200 MHz 8 bit FADCs, designed previously  for another application, to digitize preamplifier  signals directly without an intermediate shaper amplifier.  Since charge dispersion pulses are slow and signals were integrated over a few hundred ns during analysis, 25 to 40 MHz FADCs would have been adequate. 

 The data analysis  
method is similar to that used in our previous publication 
\cite{cit:tpc1b} on GEM-TPC with conventional direct charge readout except for the amplitude reconstruction technique.
For the direct charge measurement, signals result only from the 
charge deposit on a pad. Depending on the transverse diffusion in the 
TPC gas, one or more pads in a row have a signal. In this type of normal TPC readout, all signals
have the same shape, {\it e.g.} rise-time, and the maximum amplitude is proportional
to the charge collected by the pad. The pad response function ({\it PRF}) can be
evaluated from the known diffusion properties of the gas and readout geometry. 

For the charge dispersion readout, in contrast, pads away from the region of direct
 charge collection on the anode may still see  measurable
signals due to the $RC$ dispersion of the cluster charge on the resistive surface.
The observed amplitude and the charge pulse shape depends on the distance
of the pad with respect to the track ionization clusters and the characteristics of
the front-end electronics. Pads seeing a part of the direct charge on the anode will
have a prompt signal with a fast rise-time, while other pads with signals resulting only from charge
dispersion will have a smaller slower rise-time delayed signal depending on the distance to the
track. In principle, a determination of the
track {\it PRF} is possible starting from  the charge dispersion model.  
However, small {\it RC} inhomogeneities of the 
resistive anode readout structure  introduce systematic effects which make the theoretical
{\it PRF} deviate from the measurement, as observed in Reference  \cite{cit:simulation}. For the present analysis, the {\it PRF} as well as the systematic effects 
are determined empirically from the internal consistency of  a subset of cosmic ray track data used only for calibration.   The remaining part of the data is used for resolution studies.

\begin{figure}[tp]
\centerline{\mbox{\epsfxsize=\textwidth \epsffile{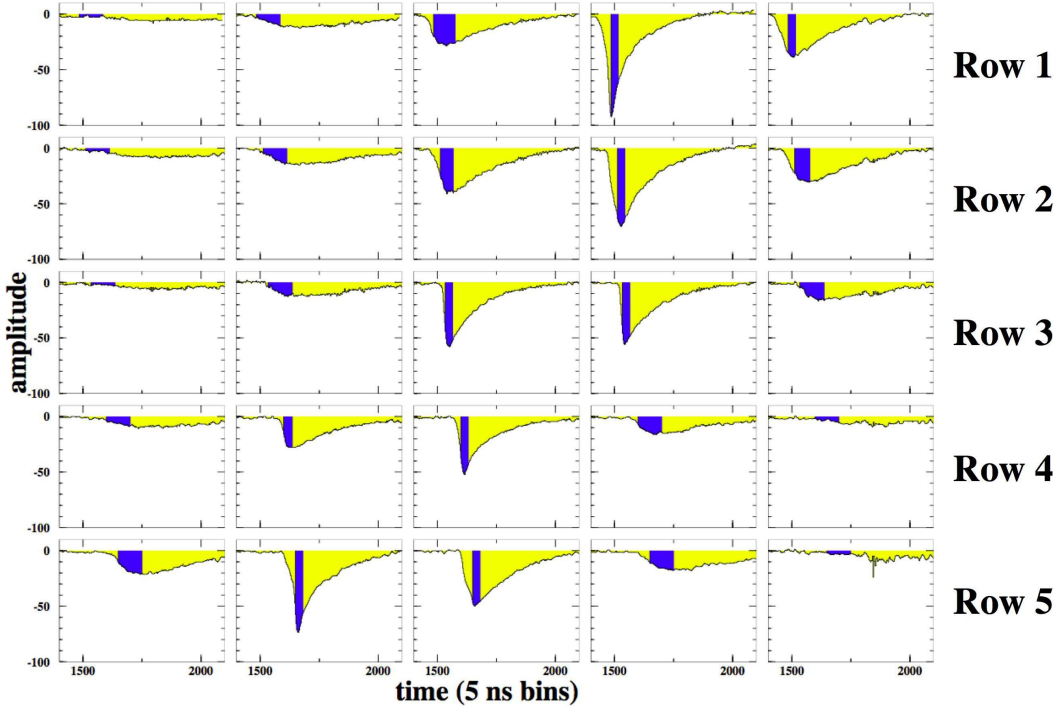}}}
\caption{\label{fig:event}
Track display plot showing observed pulse shapes for a cosmic ray track for the five rows of  2 mm x 6 mm GEM readout pads used for tracking.  The dark shaded areas indicate the regions used
to compute signal amplitudes  to determine the pad response function  ({\it PRF}), as explained in the text. The track parameters are: drift distance z = 1.97 cm,  $\phi = 0.15$ radians and $\theta = -0.70$ radians.} 
\end{figure}

As both the rise time and the pulse height carry track position information, the {\it PRF} will depend 
on the method used to reconstruct the pad signal amplitude from the measured  
pulse. The following method uses both the time and pulse height information to obtain a narrower {\it PRF}  
with shorter tails. For a given pad row, the largest pulse is identified and its amplitude
calculated by maximizing the average pulse height in a 150 ns window. 
The large fast rise time pulse arises mainly from the primary charge. For
a single track,  adjacent  pads in the row have delayed slower rising  smaller signals;
which reach their maximum peak pulse heights later.
For these pads, the start of the integration time window is kept at the value 
obtained from the largest pad signal and the width is increased to maximize
the reconstructed amplitude. The maximum window width is limited to 500 ns. 

Fast pulses, from the primary charge, are thus averaged
only over a short time period, leading to a larger calculated amplitude. Slower 
rising smaller pulses are averaged over a longer time window,
improving the signal to noise ratio. Since the start of the time window is determined by the main
pulse in the row, late pulses will be reconstructed with a smaller computed amplitude as well
leading to a suppression of the tails of the pad response function.
Figure~\ref{fig:event} shows the observed pulses for a cosmic ray track for
the five tracking rows of pads. The time bins used for the determination of the
amplitudes are also indicated. Differences in the shapes between
the main pulse dominated by primary charge and pulses from
charge dispersion on pads farther away are visible.

\begin{figure}[b]
\centerline{\mbox{\epsfxsize=11cm \epsffile{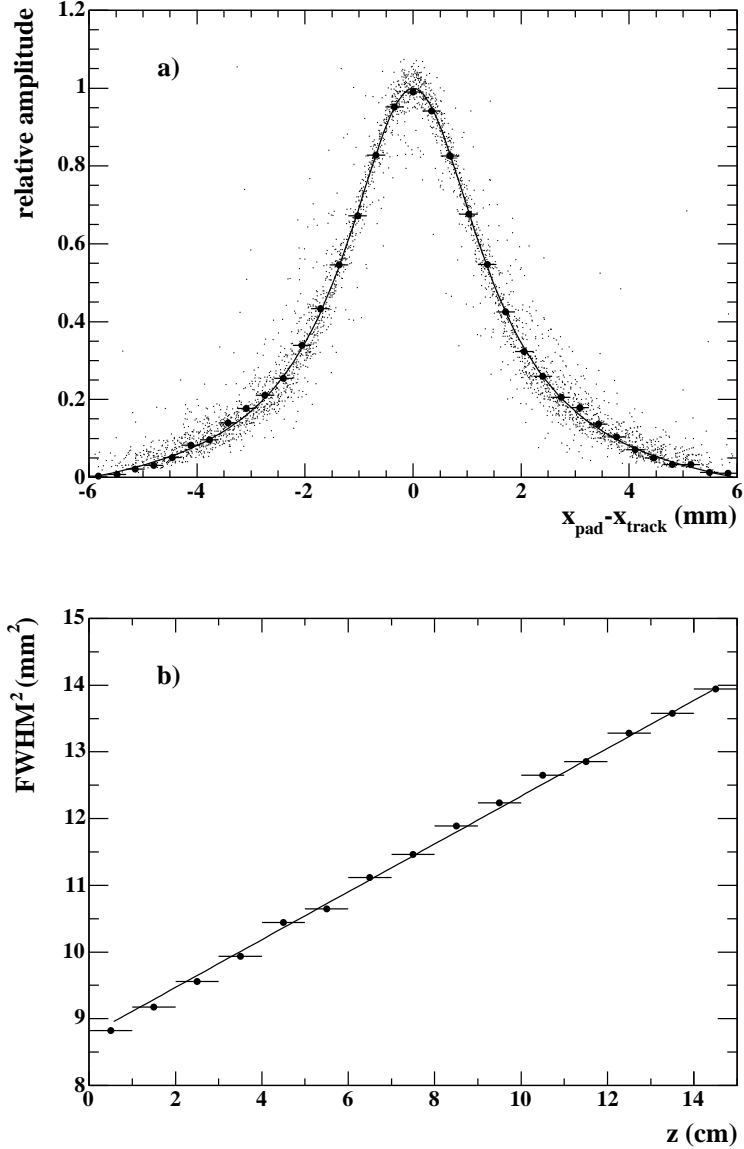}}}
\caption[]{\label{fig:prf}
a) Determination of the pad response function ({\it PRF})  from the calibration data set for the first 1 cm drift.  The  figure shows measured relative pulse amplitudes as a function of the
track x coordinate relative to pad centres and the fit to the {\it PRF} parametric form  given by Eq. \ref{eq:prf}.
 
b) The {\it PRF} as a function of drift distance was determined in 1 cm steps. The dependence of the square of the {\it FWHM} of the {\it PRF}  on the drift distance  was found to be linear.} 
\end{figure}

The  resolution study was restricted to  track angles $|\phi|<5^\circ$. The track fit of the reconstruction analysis made use of a
pad response function {\it PRF} determined  from the calibration data  set.  The {\it PRF} was determined as a function of drift distance and as mentioned before,
 the calibration data set was not used for  resolution studies.

 Figure~\ref{fig:prf}a shows the  {\it PRF}  data for  drift distances up to 1 cm. The relative 
amplitude is shown as a function of the distance between the pad-center 
and the track.  The {\it PRF} was determined in 1 cm steps and parameterized with a ratio of two symmetric 4th order polynomials:

\begin{equation}
PRF(x,\Gamma,\Delta,a,b ) = \frac{1 + a_2 x^2 + a_4 x^4}{1 + b_2 x^2 + b_4 x^4} \; 
\label{eq:prf}
\end{equation}

~~~~~with

\ba
a_2 & = & - (2/\Delta)^2 \, (1 + a) \nonumber \\
a_4 & = & (2/\Delta)^4 \, a \nonumber \\
b_2 & = &  (2/\Gamma)^2 \left(1 - b -2(1+a)\left(\frac{\Gamma}{\Delta}\right)^2 +
2a\left(\frac{\Gamma}{\Delta}\right)^4 \right) \nonumber \\
b_4 & = & (2/\Gamma)^4 \, b, \nonumber 
\ea

where in principle all parameters,  full-width-half-maximum \FWHM\ 
($\Gamma$), base width $\Delta$, and scale parameters $a$ and $b$, 
depend on the drift distance.
 For the present data set, a linear parameterization could be used for the square of  \FWHM\,  as shown in Figure~\ref{fig:prf}b.  The other parameters at
 $b=0$, $a=-0.3$ and $\Delta=11.9$ mm were held constant.
 
Since the track fit uses a $\chi^2$ minimization,  the amplitude measurement errors must also be determined from the
data. In our case, this error is dominated by systematic effects leading to
a mainly linear dependence on the amplitude.

\section{Analysis and  results}
\label{sec:results}

\begin{figure}[tp]
\centerline{\mbox{\epsfxsize=10cm \epsffile{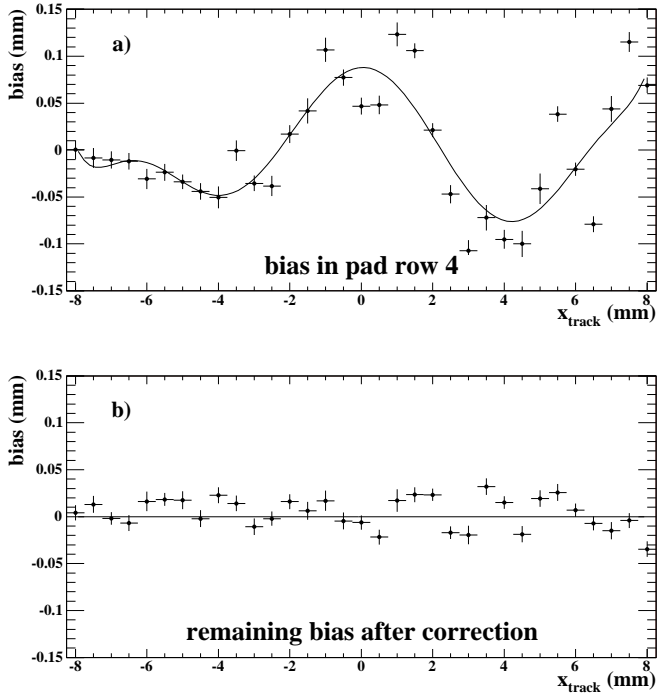}}}
\caption[]{\label{fig:bias}
Bias in the determination of track position before and after correction. The figure shows position residuals ($x_{\rm row}-x_{\rm track}$) for row 4 (see Fig \ref{fig:event}) as a function of $x_{\rm track}$, 
a) before and b) after bias correction.} 
\end{figure}

As in our previous paper \cite{cit:tpc1b}, the track parameters 
$x_0$ and $\phi$ are
determined from a global fit to all pad amplitudes of a given event. 
We use a right-handed coordinate system with the $x$-coordinate
horizontal and the $y$-coordinate parallel to the pad length; the 
$z$-coordinate corresponds to the drift distance with $z=0$ at the
first GEM stage. The azimuthal angle $\phi$ and the polar angle $\theta$
are measured with respect to the $y$-axis. The position in a row
$x_{\rm row}$ is determined from a separate one-parameter track fit 
to this
row only using the known track angle $\phi$. Figure~\ref{fig:bias}a
shows the mean of the track residuals $x_{\rm row}-x_{\rm track}$ for row 4  (see Fig \ref{fig:event})   as a function of
$x_{\rm track} = x_0 + \tan{\phi}*y_{\rm row}$, where $y_{\rm row}$ is
the $y$ position of the row. A bias of up to 130 \mum\ is observed
which we attribute to small local variations in the $RC$ 
from imperfections in the quality of materials and technique used presently in laminating the resistive anode readout assembly.
 The bias is intrinsic to the detector and does not change with time. It can therefore be easily corrected. 
The calibration data set used for the {\it PRF} determination is also used to determine the bias correction
for each pad row in 500 $\mu$m steps. Figure~\ref{fig:bias}b shows the mean track residuals for the central pad row after bias correction. The remaining bias after correction was small.

\begin{figure}[bp]
\centerline{\mbox{\epsfxsize=9cm \epsffile{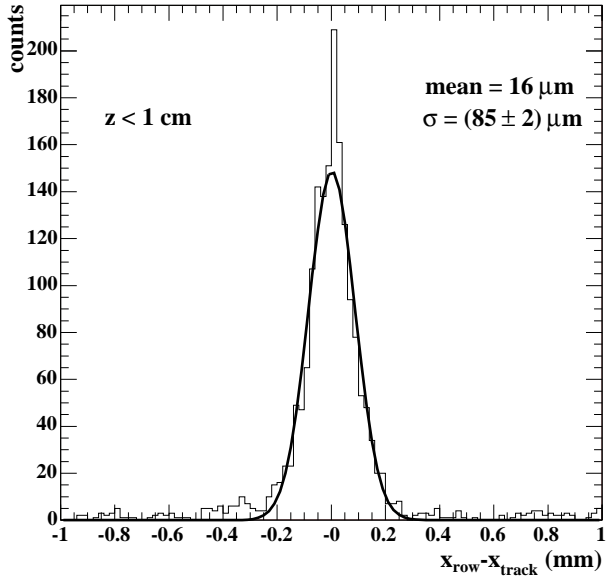}}}
\caption[]{\label{fig:residual}
Position residuals $x_{\rm row}-x_{\rm track}$ for short drift
distance $z < 1 \mbox{ cm}$ and track angles $|\phi|<5^\circ$ after 
bias correction. The mean corresponds to the remaining bias.} 
\end{figure}

\begin{figure}[t]
\centerline{\mbox{\epsfxsize=10cm \epsffile{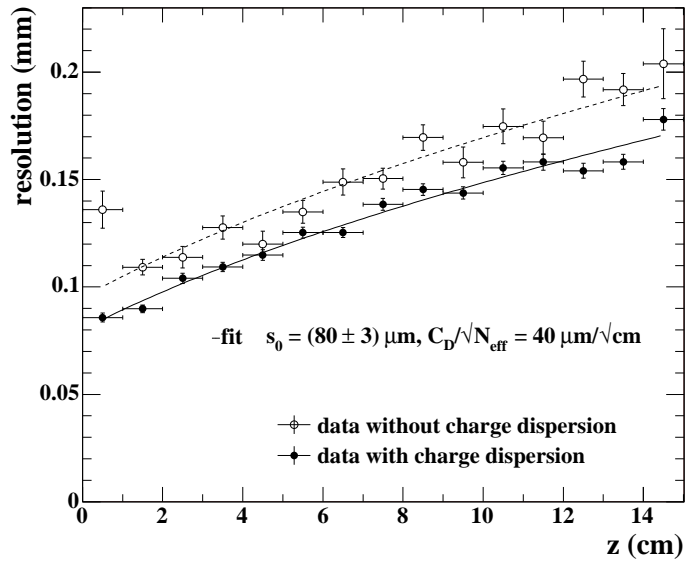}}}
\caption[]{\label{fig:resolution}
Transverse resolution for track angles $|\phi|<5^\circ$ as a function of 
drift distance $z$ for 2 mm wide pads. The data with charge dispersion is fitted to the resolution expected from diffusion in the TPC gas and electron statistics (Eq. \ref{eq:reso}) (solid line).
For comparison, the GEM-TPC resolution with direct charge 
readout from our previous work \cite{cit:tpc1b} is also shown (dashed line).} 
\end{figure}

Figure~\ref{fig:residual} shows the distribution of the residuals
for tracks with $|\phi| < 5^\circ$ and small drift distance
$z <  1$ cm. 
As in our previous publication \cite{cit:tpc1b} the resolution is given by
the geometric mean of  standard deviations of residuals from track fits done in two different ways:  including and excluding the row for which the resolution is being determined. The measured resolution as a function of drift distance
is shown in Figure~\ref{fig:resolution} together with a fit to the function:

\begin{equation}
  s = \sqrt{ s_0^2 + \frac{C_{\rm D}^2 z}{N_{\rm eff}}} ~~,
  \label{eq:reso}
\end{equation}


where $s_0$ is the resolution at $z=0$, $C_D$ is the transverse diffusion constant and 
$N_{\rm eff}$ is the effective number of electrons along the track in a row.

Electronic noise and systematic effects contribute to the constant term $s_0$, the resolution at zero drift distance. The constant term $s_0$  is about 80 $ \mu$m for the  
charge dispersion readout. In contrast, as shown in Figure 5, the TPC resolution with the conventional GEM readout for drift distances approaching zero would be much larger (138 $ \mu$m at 5 mm), due to lack of precision in pad centroid determination from diffusion. The resolution for the conventional GEM readout improves with increasing transverse diffusion for larger drift distances.  Nevertheless, 
  the TPC resolution obtained with the charge dispersion  
readout remains better than with the conventional GEM readout \cite{cit:tpc1b} even 
for larger drift distances. This is due to the fact that the charge dispersion phenomena can be completely described by material properties and geometry and  the centroid of the dispersed charge signals on the resistive anode can be accurately determined,  in contrast to centroid determination from diffusion, which is statistical in nature.


\section{Summary and outlook}

A GEM-TPC with a charge dispersion readout system incorporating a resistive anode has been
used to measure particle track resolutions for the first time. 
The resistive anode allows a controlled dispersion of the track charge clusters
over several pads which can be used for a precise determination of  the charge centroid.
Using 2 mm x 6 mm pads, we have shown that charge dispersion  improves the GEM-TPC
resolution significantly over that achievable with conventional direct charge readout,  both at short 
and at long drift distances. 
Imperfections in the resistive anode assembly and materials lead to a position measurement bias which can be easily corrected. The bias remaining after correction is small.  With improvements in  fabrication techniques and the quality of materials, the measurement bias will be reduced further.  The TPC pad readout signals were digitized at 200 MHz for the results reported here.  We are in the process of developing slower 25 to 40 MHz digitizers which will be adequate  for these type of measurements.

\section*{Acknowledgments}
We thank Bob Carnegie and Hans Mes for numerous discussions and helpful suggestions throughout the course of this work. 
Our TPC front-end charge pre-amplifiers were used previously for the ALEPH TPC readout at LEP and we thank Ron Settles for providing these to us. 
Ernie Neuheimer was our electronics expert for the project and he  designed, built and did much of the troubleshooting of  the front-end and readout electronics. Our
mechanical engineers, Morley O'Neill initially and Vance Strickland subsequently, worked on designing
the TPC assembly and developing the clean-room facility used for 
the detector assembly. Philippe Gravelle was always willing and available to help us
solve a variety of technical problems. Our CO-OP students Alasdair Rankin, Steven Kennedy,
Roberta Kelly and David Jack worked on the commissioning of the detector as well as writing parts of the data acquisition and analysis software.   Finally, we thank Alain Bellerive for a critical reading of the manuscript and for identifying parts that needed improvements. This research was supported by a project grant from the Natural Sciences and Engineering Research Council of Canada. TRIUMF receives federal funding via a contribution agreement through the National Research Council of Canada.

\bibliographystyle{elsart-num}
\bibliography{pr}
\end{document}